\documentclass[]{interact}

\usepackage{epstopdf}                              % To incorporate .eps illustrations using PDFLaTeX, etc.
\usepackage{subfigure}                             % Support for small, `sub' figures and tables

\usepackage[numbers,sort&compress,merge]{natbib}   % Citation support using natbib.sty
\bibpunct[, ]{(}{)}{,}{n}{,}{,}                    % Citation support using natbib.sty
\setcitestyle{square,comma}                        % Force citations to square brackets - why doesn't preceding line do this?
\renewcommand\bibfont{\fontsize{10}{12}\selectfont}% Bibliography support using natbib.sty
\renewcommand\citenumfont[1]{\textit{#1}}          % Citation numbers in italic font using natbib.sty
\renewcommand\bibnumfmt[1]{(#1)}                   % Parentheses enclose ref. numbers in list using natbib.sty
\DeclareMathOperator\erf{erf}                      % Add error function to list of operators

\theoremstyle{plain}                               % Theorem-like structures

\theoremstyle{definition}

\theoremstyle{remark}

\begin{document}

\articletype{SPECIAL ISSUE ARTICLE}

\title{Matterwave interferometric velocimetry of cold Rb atoms}

\author{
\name{Max Carey\textsuperscript{a}, Mohammad Belal\textsuperscript{a}, Matthew Himsworth\textsuperscript{a}, James Bateman\textsuperscript{b} and Tim Freegarde\textsuperscript{a}\thanks{CONTACT Tim Freegarde. Email: tim.freegarde@soton.ac.uk}}
\affil{\textsuperscript{a}School of Physics \& Astronomy, University of Southampton, Southampton SO17~1BJ, UK; \textsuperscript{b}Department of Physics, Swansea University, Swansea SA2~8PP, UK}
}
\makeatletter
\gdef\@received{Compiled October 10, 2017}
\makeatother
\maketitle

\begin{abstract}
We consider the matterwave interferometric measurement of atomic velocities, which forms a building block for all matterwave inertial measurements. A theoretical analysis, addressing both the laboratory and atomic frames and accounting for residual Doppler sensitivity in the beamsplitter and recombiner pulses, is followed by an experimental demonstration, with measurements of the velocity distribution within a 20~$\umu$K cloud of rubidium atoms. Our experiments use Raman transitions between the long-lived ground hyperfine states, and allow quadrature measurements that yield the full complex interferometer signal and hence discriminate between positive and negative velocities. The technique is most suitable for measurement of colder samples.
\end{abstract}

\begin{keywords}
Velocity measurement; atom interferometry; cold atoms; laser cooling; ion traps
\end{keywords}

\section{Introduction}
For vapour phase atoms to reveal their quantum-mechanical characteristics, they must usually be cooled. At ultracold temperatures below 1~mK, reduced translational motion means that atomic collisions are rare, atoms remain within an experimental region for long enough to be manipulated and observed, Doppler shifts do not mask more subtle phenomena, and trapped species are strongly localized. The quantum state coherence is then largely unperturbed, energy levels are well defined and their spectra simplified, and the kinetic energy available for collisional exchange is miniscule. Reduced to a small set of better-defined, longer-lived quantum states, cold atoms and ions allow classic manifestations of quantum statistics -- Bose-Einstein condensation, Mott insulator and Dicke phase transitions -- and are the basis for a plethora of information processing and sensing mechanisms.

Although the sensitivities of quantum superpositions to accelerations, rotations and gravitational fields and gradients have been widely studied~\cite{Berman96,Miffre06}, there have been few investigations of the velocimetry process that lies at their hearts. This is perhaps because it cannot be used as a sensor of the apparatus' velocity, since the atom cloud that forms the test mass begins in the same inertial frame as the apparatus. Weitz and H\"{a}nsch proposed the use of velocity-dependent atom interferometry for frequency-independent laser cooling~\cite{Weitz00}, subsequently demonstrated by Dunning {\em et al.}~\cite{Dunning15}; and Weiss {\em et al.} used the technique for measurement of the photon recoil~\cite{Weiss93}. Separately, Shirley~\cite{Shirley97} used Fourier analysis to obtain velocity distributions from Ramsey lineshapes, but with no velocity sensitivity in the interaction or wavefunction evolution the Ramsey method was effectively an alternative modulation source for Fourier transform time-of-flight velocimetry~\cite{Knorr86}.

In this paper, we show that atom interferometry provides a useful tool for velocity measurement within atom clouds themselves. The process is the building block for all inertial sensing matterwave interferometers, which are effectively differential pairs of the Ramsey sequences addressed here. We therefore begin with an analysis of the fundamental principles, relating different perspectives and extracting some key results. We then describe our experimental investigation, and present results for cold Rb atoms with a temperature around 20~$\umu$K. With limited laser intensities, we observe a residual Doppler sensitivity in our beamsplitter pulses, which limits the resolution of our measurements: a theoretical analysis of this effect forms the Appendix.

\section{Interferometric velocimetry}
According to the frame of reference, velocity plays different roles in matterwave interferometry, defining both the particle's classical trajectory and its quantum mechanical evolution. We examine the interferometric process from both the laboratory and the atomic rest frames, and obtain common results for the velocity dependence of the interferometer phase.

\subsection{Laboratory frame}
Atom interferometry is commonly depicted in the laboratory frame, in which the apparatus is fixed and the atoms move. Interaction with a laser forms and resolves a quantum mechanical superposition, and the interferometer reveals the residual energy difference, after subtraction of the laser frequency, between the two superposed states. The laser thus provides both a frequency reference for the apparatus, and the $\upi/2$ and $\upi$-pulses that act as the matterwave beamsplitters and mirrors.

\begin{figure}
\centering
%\subfigure[xx.]{
%\resizebox*{5cm}{!}{\includegraphics{./figures/f1803a}}}\hspace{5pt}
%\subfigure[xx.]{
%\resizebox*{5cm}{!}{\hspace{5mm}}}
\includegraphics[width=45mm]{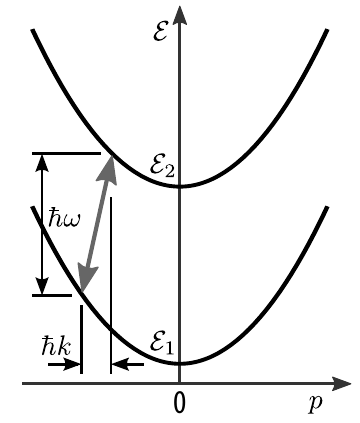}
\caption{Graphical representation of the conservation of energy $\mathcal{E}$ and momentum $p$ during photon absorption or emission. The photon is represented by the line of slope $c$, whose length varies with the atomic momentum in accordance with the classical Doppler shift.}
\label{energies}
\end{figure}

The sensitivity to velocity is apparent when the kinetic energy, and its modification by the photon recoil, is included in the atomic Lagrangian~\cite{Gustavson00}. Although many atom interferometers use two-photon Raman transitions, the principles are more simply demonstrated by an atom with two electronic states $\vert 1 \rangle$ and $\vert 2 \rangle$ that may be radiatively coupled by absorption or emission of a single photon of frequency $\omega$ and wavevector $\mathbf{k}$. If the electronic energies of the states are $\mathcal{E}_{1,2}$, and we write the full electronic+motional states as $\vert 1, \bf{p}_{1} \rangle$ and $\vert 2, \bf{p}_{2} \rangle$, then conservation of energy and momentum requires
\begin{eqnarray}
\hbar \omega & = & \left( \mathcal{E}_{2} + \frac{\vert\mathbf{p}_{2}\vert^{2}}{2 m} \right) - \left(\mathcal{E}_{1} + \frac{\vert\mathbf{p}_{1}\vert^{2}}{2 m} \label{energy} \right) \\
\hbar \mathbf{k} & = & \mathbf{p}_{2} - \mathbf{p}_{1}, \label{momentum}
\end{eqnarray}
where $m$ is the atom's mass. When we write $\mathbf{p}_{1,2} \equiv \mathbf{p} \mp \frac{1}{2}\hbar\mathbf{k}$~\cite{Barnett10}, so that the photon couples states $\vert 1, \mathbf{p}\!-\!\frac{1}{2}\hbar \mathbf{k} \rangle$ and $\vert 2, \mathbf{p}\!+\!\frac{1}{2}\hbar \mathbf{k} \rangle$ to satisfy Equation~(\ref{momentum}), Equation~(\ref{energy}) becomes
\begin{equation}
\hbar \Delta = \hbar\omega - \left( \mathcal{E}_{2}-\mathcal{E}_{1} \right) = \frac{\mathbf{p}\cdot\hbar\mathbf{k}}{m},
\end{equation}
where $\Delta$ is the detuning from resonance. This is simply the classical Doppler shift, and is illustrated graphically in Figure~\ref{energies}(a), in which the photon is represented by a line of slope $c$ which must connect the two parabolas, and whose length therefore increases with the component of the atomic momentum in the direction of photon propagation.

If the atom is not subject to any external field between the pulsed interactions that form the interferometer beamsplitter and recombiner at $t\!=\!t_{1,2}$, the atomic Lagrangian during this period will be
\begin{equation}
\mathcal{L} = \frac{\vert \mathbf{p} \vert^{2}}{2m}
\label{lagrangian}
\end{equation}
so that
\begin{equation}
\mathcal{L}_{2} - \mathcal{L}_{1} = \hbar\Delta.
\end{equation}
The interferometer phase $\varphi$ is then simply~\cite{Feynman48}
\begin{eqnarray}
\varphi & = & \frac{1}{\hbar} \int_{t_{1}}^{t_{2}} \left( \mathcal{L}_{2} - \mathcal{L}_{1} \right) \mathrm{d}t \label{pathintegral} \nonumber \\
& = & \int_{t_{1}}^{t_{2}} \Delta \, \mathrm{d}t \nonumber \\
& = & \frac{\mathbf{p}\cdot\mathbf{k}}{m} T = \mathbf{k} \cdot \mathbf{v} T, \label{phase}
\end{eqnarray}
where $T \equiv t_{2}-t_{1}$ is the interferometer measurement period.

\begin{figure}
\centering
\includegraphics[width=100mm]{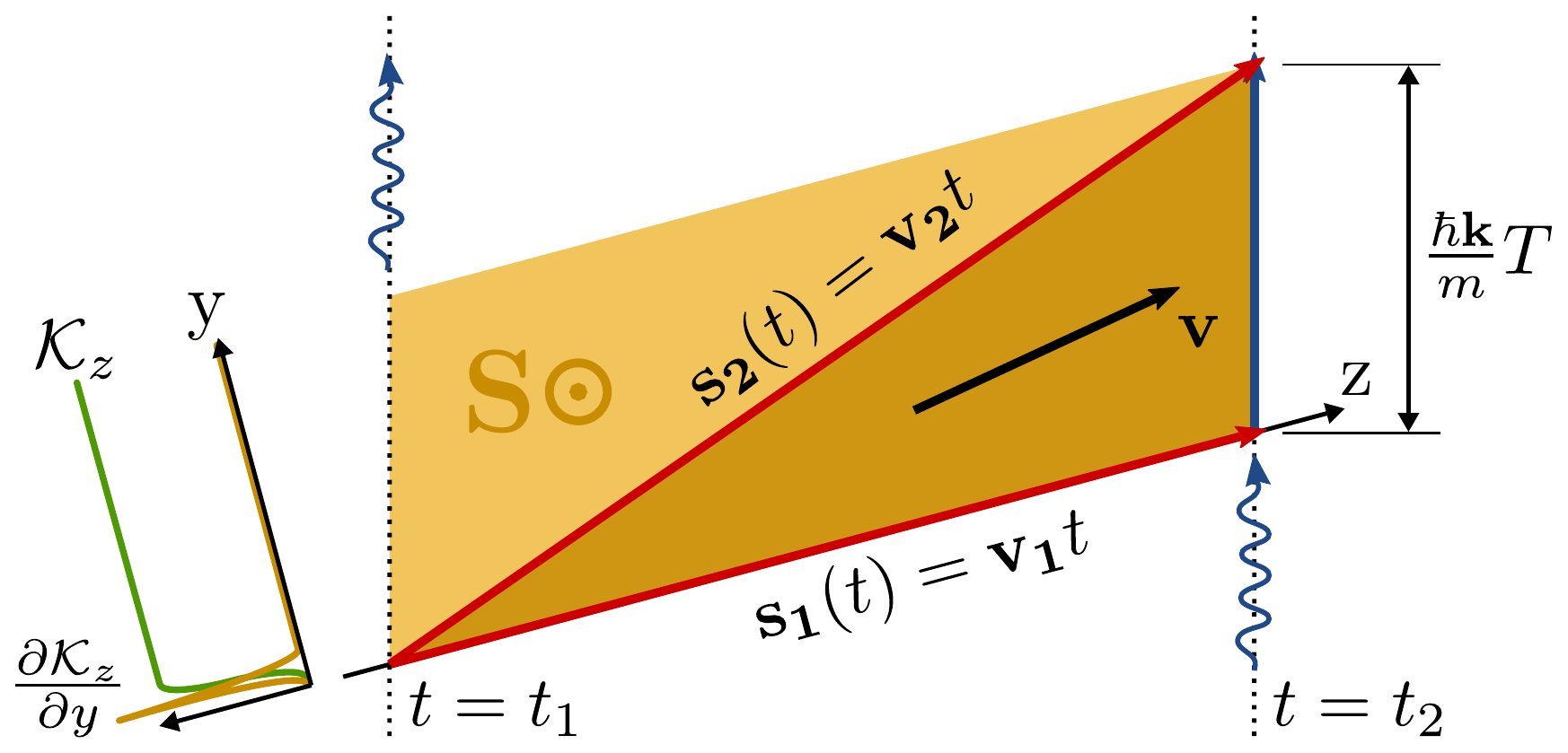}
\caption{Interferometer paths $\mathbf{s}_{1,2}(t)$ and area $\mathbf{S}$ used for calculating the path integral of the quantum mechanical action.}
\label{paths}
\end{figure}

For an alternative derivation of Equation~(\ref{phase}), we define
\begin{equation}
\mathcal{L} = \bm{\mathcal{K}}\cdot\mathbf{p}
\end{equation}
so that the path integral~\cite{Storey94} of Equation~(\ref{pathintegral}) may be re-written as
% see Storey p2022 for use of Stokes theorem
\begin{eqnarray}
\varphi & = & \frac{1}{\hbar} \left( \int_{t_{1}}^{t_{2}} \bm{\mathcal{K}}\cdot\mathbf{p}_{2} \, \mathrm{d}t - \int_{t_{1}}^{t_{2}} \bm{\mathcal{K}}\cdot\mathbf{p}_{1} \, \mathrm{d}t \right) \nonumber \\
& = & \frac{m}{\hbar} \left( \int_{\mathbf{s}_{2}(t_{1})}^{\mathbf{s}_{2}(t_{2})} \bm{\mathcal{K}}_{2}\cdot\mathrm{d}\mathbf{s}_{2} - \int_{\mathbf{s}_{1}(t_{1})}^{\mathbf{s}_{1}(t_{2})} \bm{\mathcal{K}}_{1}\cdot\mathrm{d}\mathbf{s}_{1} \right) \nonumber \\
& = & \frac{m}{\hbar} \left( \oint \bm{\mathcal{K}}\cdot\mathrm{d}\mathbf{s} + \int_{\mathbf{s}_{1}(t_{2})}^{\mathbf{s}_{2}(t_{2})} \bm{\mathcal{K}}\cdot\mathrm{d}\mathbf{s} \right)  \equiv \varphi_{\mathrm{loop}} + \varphi_{12} \label{phase2}
\end{eqnarray}
where $\mathbf{s}_{1,2}$ is the path followed by state $(1,2)$ from $t=t_{1}$ to $t=t_{2}$, as shown in Figure~\ref{paths}, and the final term $\varphi_{12}$ is the shift due to the separation of the wavepackets when they are recombined~\cite{Bongs06}.

The Kelvin-Stokes theorem then allows the first term in Equation~(\ref{phase2}) to be re-written as
\begin{equation}
\varphi_{\mathrm{loop}} = \frac{m}{\hbar} \iint \left( \bm{\nabla \times} \bm{\mathcal{K}} \right) \cdot \mathrm{d}\mathbf{S},
\end{equation}
where $\mathbf{S}$ is the area enclosed by the interferometer. When the interferometer is used to sense an external magnetic, electric or gravitational field, or equivalent non-inertial motion, this will be represented by $\bm{\nabla \times} \bm{\mathcal{K}}$ and the interferometer sensitivity will commonly scale with the enclosed area.

For the Lagrangian of Equation~(\ref{lagrangian}), $\bm{\mathcal{K}} = \mathbf{p}/2 m = \mathbf{v}/2$ and, in the absence of external fields, is uniform except during the beamsplitter interaction. To cast this into the form above, we consider the impulse to be extended to occur over a finite time and distance. Writing $\mathbf{z} \equiv \mathbf{v}_{1}t$, and defining the $y$ axis to lie in the $\mathbf{k}$-$\mathbf{v}$ plane, we may write the first term in the interferometer phase as
\begin{equation}
\varphi_{\mathrm{loop}} = \frac{m}{\hbar} \iint \left( \bm{\nabla \times \mathcal{K}} \right) \cdot \bm{\hat{\mathrm{x}}} \, \mathrm{d}y\mathrm{d}z  = \frac{m}{\hbar} \iint \left( \frac{\partial \mathcal{K}_{z}}{\partial y} -\frac{\partial \mathcal{K}_{y}}{\partial z}  \right) \mathrm{d}y\mathrm{d}z,
\end{equation}
where the integral is over the triangular area enclosed by the interferometer paths. Since the beamsplitter impulse is assumed to occur at around $t=0$, $\partial \mathcal{K}_{z}/\partial y$ will be zero in most of the triangle, and the area can be extended to a trapezium without affecting the result, allowing the integrals over $y$ and $z$ to be separated.

For the initial state $\vert 1, \mathbf{p}\!-\!\hbar \mathbf{k}/2 \rangle$ to receive no impulse, $\partial \bm{\mathcal{K}}/\partial z = 0$, while
\begin{equation}
\int \frac{\partial \mathcal{K}_{z}}{\partial y} \mathrm{d}y = \frac{\hbar k_{z}}{2 m} = \frac{\hbar \mathbf{k}\cdot\mathbf{v}_{1}}{2 m v_{1}}.
\end{equation}
The first term in Equation~(\ref{phase2}) hence becomes
\begin{equation}
\varphi_{\mathrm{loop}} = \frac{m}{\hbar} \int_{0}^{v_{1}T} \frac{\hbar \mathbf{k}\cdot\mathbf{v}_{1}}{2 m v_{1}} \mathrm{d}z = \frac{1}{2} \mathbf{k}\cdot\mathbf{v}_{1} T,
\label{phase3}
\end{equation}

Since the path separation at $t=t_{2}$ will be $\hbar \mathbf{k}T/m$, the second term in Equation~(\ref{phase2}) may be written, to first order in $\hbar \mathbf{k}/m$, as
\begin{equation}
\varphi_{12} = \frac{m}{\hbar} \int_{\mathbf{s}_{1}(t_{2})}^{\mathbf{s}_{1}(t_{2})+ \hbar \mathbf{k} T/m} \frac{\mathbf{v}_{2}}{2}\cdot\mathrm{d}\mathbf{s} = \frac{m}{\hbar} \frac{\mathbf{v}_{2}}{2} \cdot \frac{\hbar \mathbf{k} T}{m} = \frac{1}{2} \mathbf{k}\cdot\mathbf{v}_{2} T,
\label{phase4}
\end{equation}
so that the total interferometer phase, the sum of Equations~(\ref{phase3}) and (\ref{phase4}), will be
\begin{equation}
\varphi = \mathbf{k}\cdot\mathbf{v} T,
\label{phase5}
\end{equation}
reproducing Equation~(\ref{phase}).

\subsection{Atomic frame}
In its inertial frame, the atom is a precise clock, set by the beamsplitter interaction and subsequently read by comparing it with the phase of the recombiner. Any relative change in the optical field phase at the atom, due either to variations of laser phase or frequency or to movement of the apparatus with respect to the atom, shifts the interferometer signal. From this perspective, atom interferometric inertial measurement is a microscopic version of the traditional method of determining longitude by measuring the phase of a ship's clock,  synchronized to noon at the meridian, relative to the periodic variations in the sun's elevation above the horizon~\cite{Sobel95}. The change in phase and hence position during a given measurement time reveals the clock's velocity~\cite{Preclikova14}.

For the interferometer phase to reflect the displacement $\mathbf{\delta r} \equiv \mathbf{v}T$, we require the phase $\vartheta$ of the optical field with respect to the atomic clock to vary linearly with position in the measurement direction, i.e., for all $\mathbf{v}$ and $T$,
\begin{equation}
\vartheta(\mathbf{r}_{0}\!+\!\mathbf{\delta r},t_{2})-\vartheta(\mathbf{r}_{0},t_{1}) = \mathbf{k}\cdot\mathbf{\delta r} ,
\end{equation}
where the beamsplitter interaction synchronizes the atomic oscillator at position $\mathbf{r}_{0}$ at time $t_{1}$ and the atom is interrogated by the recombiner interaction at time $t_{2} = t_{1} + T$. Although $\mathbf{k}$ may here be regarded as an arbitrary vector constant, it will indeed prove to be the field wavevector previously defined. We hence obtain
\begin{equation}
\bm{\nabla} \vartheta \cdot \mathbf{\delta r} + \frac{\partial \vartheta}{\partial t} T = \mathbf{k} \cdot \mathbf{\delta r},
\end{equation}
so $\partial \vartheta/\partial t = 0$ and $\bm{\nabla} \vartheta = \mathbf{k}$, from which we determine that the optical phase must at any point track the atomic phase $\omega_{0} t$ and depend spatially upon $\mathbf{k}\cdot\mathbf{r}$. The phase of the optical field must thus have the form $(\mathbf{k}\cdot\mathbf{r}\!-\!\omega_{0}t)$ characteristic of a travelling plane wave. The rate of variation in optical phase at the position of the atom is again simply the Doppler shift; while we have here considered Galilean transformation between the apparatus and atomic frames, equivalent results may be obtained by relativistic Lorentz transformation~\cite{Barnett10}.

%\begin{equation}
%\frac{\vartheta(\mathbf{r}_{0} \! + \! \mathbf{v}T,t_{1}\!+\!T)-\vartheta(\mathbf{r}_{0} \! + \! \mathbf{v}T,t_{1})}{T} + \frac{\vartheta(\mathbf{r}_{0} \! + \! %\mathbf{v}T,t_{1})-\vartheta(\mathbf{r}_{0},t_{1})}{T} = \mathbf{v}\cdot\mathbf{k} .
%\end{equation}

Interaction with the optical field imparts an impulse to the atom classically through the Lorentz force upon the dipole induced by the electric field, in the presence of the magnetic field which follows, via Amp\`{e}re's law, from the spatial variation of the electric field~\cite{FeynmanPhoton,Hinds09}. Quantum mechanically, the impulse excites a two-state superposition whose phase varies spatially with that of the optical field as $\mathbf{k}\cdot\mathbf{r}$, thus giving the excited state an impulse $\hbar \mathbf{k}$. We note that if the atomic centre-of-mass wavefunction is localized to within an optical wavelength, the momentum uncertainty will exceed the single photon impulse.

\section{Interferometric velocimetry}
The essential stages of an atom interferometer are a source of 2-level atoms, a means of preparation into one of the two states $\vert 1 \rangle$, a $\upi/2$ or beamsplitter interaction that implements a $\upi/2$ rotation on the Bloch sphere~\cite{Feynman57} to leave the atoms in a equal superposition, a period of free evolution in which to accrue the measurement phase, a further $\upi/2$ pulse to recombine the superposition, and a read-out mechanism to collapse the atoms into the two states and determine their relative population~\cite{Cronin07}.

Since the rotation performed by the recombiner interaction maps the Bloch sphere longitude onto the latitude, the interferometer signal, characterized for example by the fraction $\vert c_{2} \vert^{2}$ of the population in state $\vert 2 \rangle$, follows a sinusoidal form
\begin{equation}
\vert c_{2} \vert ^{2} = \frac{1+\cos \varphi}{2} = \frac{1}{2} \left[ 1 + \cos \left( \mathbf{k} \cdot \mathbf{v}T \right) \right] = \frac{1}{2} \left[ 1 + \cos \left( kv_{\mathrm{k}}T \right) \right],
\end{equation}
where $\varphi$ is the interferometer phase discussed above and $v_{\mathrm{k}} \equiv \mathbf{v} \cdot \mathbf{\hat{k}}$. If the interferometer signal is recorded for a range of values of $T$, each velocity class will contribute sinusoidal fringes according to the number of atoms with a given velocity component in the direction of the optical wavevector, and the total signal $\mathcal{C}(T)$ will be
\begin{equation}
\mathcal{C}(T) \propto \int \rho(\mathbf{v}) \vert c_{2} \vert^{2} \mathrm{d}\mathbf{v} = \frac{1}{2}\int \rho(v_{\mathrm{k}}) \left[ 1 + \cos \left( kv_{\mathrm{k}}T \right) \right] \mathrm{d}v_{\mathrm{k}} ,
\label{signal}
\end{equation}
where $\rho(v_{\mathrm{k}})$ is the atomic number density as a function of the velocity component along $\mathbf{\hat{k}}$. This velocity distribution can be revealed by computing the Fourier transform of Equation~(\ref{signal}) -- that is, except when $v'_{\mathrm{k}} \approx 0$,
\begin{equation}
\rho(v'_{\mathrm{k}}) \propto \int \mathcal{C}(T) \cos \left( kv'_{\mathrm{k}}T \right) \mathrm{d}T .
\end{equation}

The same fringes are hence obtained for both signs of $v'_{\mathrm{k}}$. This ambiguity may be resolved by repeating the measurement with a $\upi/2$ phase shift introduced into the optical field between the beamsplitter and recombiner yielding the signal
\begin{equation}
\mathcal{S}(T) \propto \frac{1}{2}\int \rho(v_{\mathrm{k}}) \left[ 1 + \sin \left( kv_{\mathrm{k}}T \right) \right] \mathrm{d}v_{\mathrm{k}} .
\end{equation}
The velocity distribution is then given by
\begin{equation}
\rho(v'_{\mathrm{k}}) \propto \int \left[ \mathcal{C}(T) \!-\! \mathrm{i}\,\mathcal{S}(T) \right] \exp \left( \mathrm{i}kv'_{\mathrm{k}}T \right) \mathrm{d}T
\end{equation}

While in principle the $\upi/2$ interactions are performed quickly, in practice the available laser power, spread over an area sufficient to illuminate the atom cloud with roughly uniform intensity, may be insufficient to avoid incurring some Doppler sensitivity. As discussed in the Appendix, this causes the fringes for a given velocity class to be modified in magnitude and phase. The transformed fringe signal must therefore be corrected to
\begin{equation}
\rho(v'_{\mathrm{k}}) \propto \frac{1}{\gamma(v'_{\mathrm{k}})} \int \left[ \mathcal{C}(T) \!-\! \mathrm{i}\,\mathcal{S}(T) \right] \exp \left( \mathrm{i}kv'_{\mathrm{k}}T \right) \mathrm{d}T
\end{equation}
where the factor $\gamma(v_{\mathrm{k}})$ is given in Equation~(\ref{gamma}).

In both c.w. and Ramsey spectroscopy, the signal observed is the convolution of the Doppler-shifted resonance with the cross-correlation of the atom-laser coherence. In conventional spectroscopy, the atom-laser interaction is dominated by the atomic and laser linewidths, collisions, and inhomogeneities in intensity, magnetic field and Zeeman sub-state, most of which contribute to a Voigt profile. Here, it is instead dominated by the double pulse of the Ramsey interaction, whose Fourier transform results in the sinusoidal fringes. In principle, there should be no Doppler sensitivity within the $\upi$-pulses - although power constraints mean that in our case there are, as addressed in the Appendix.

We note that, if quickly moving atoms leave the experimental region between the beamsplitter and recombiner pulses, the corresponding fringes will diminish with increasing $T$, and the derived velocity distribution hence broadened by convolution with a velocity-dependent function. Unlike conventional c.w. spectroscopy, however, measurements of velocities in the wings of the distribution are not distorted by weak interactions with more numerous atoms~\cite{Hughes17}.

\section{Experiment}
Our experimental approach resembles that previously reported~\cite{Dunning15}. $^{85}$Rb atoms are trapped and cooled in a 3D magneto-optical trap (MOT), the magnetic field gradient is turned off, and the beam intensities linearly reduced over 5~ms. Sub-Doppler cooling for 6~ms in the 3D molasses then cools the atom cloud to around $20\,\umu\mathrm{K}$.

The MOT repumping laser, resonant with the $5S_{1/2}\,F=2\rightarrow5P_{3/2}\,F=3$ transition, is then extinguished, and the atoms are optically pumped in 4~ms into the $5S_{1/2}\,F=2$ ground hyperfine state by the MOT cooling laser, which is red-detuned from the $5S_{1/2}\,F=3\rightarrow5P_{3/2}\,F=4$ transition. Three mutually orthogonal sets of shim coils cancel the residual magnetic field at the cloud position, such that the Zeeman sub-levels $m_{F}=-F\ldots F$ for each hyperfine state are degenerate to much less than the Rabi frequency $\Omega_{\mathrm{eff}}\approx 2\upi \times 350$~kHz observed for the Raman transition.

\begin{figure}
\centering
\includegraphics[width=140mm]{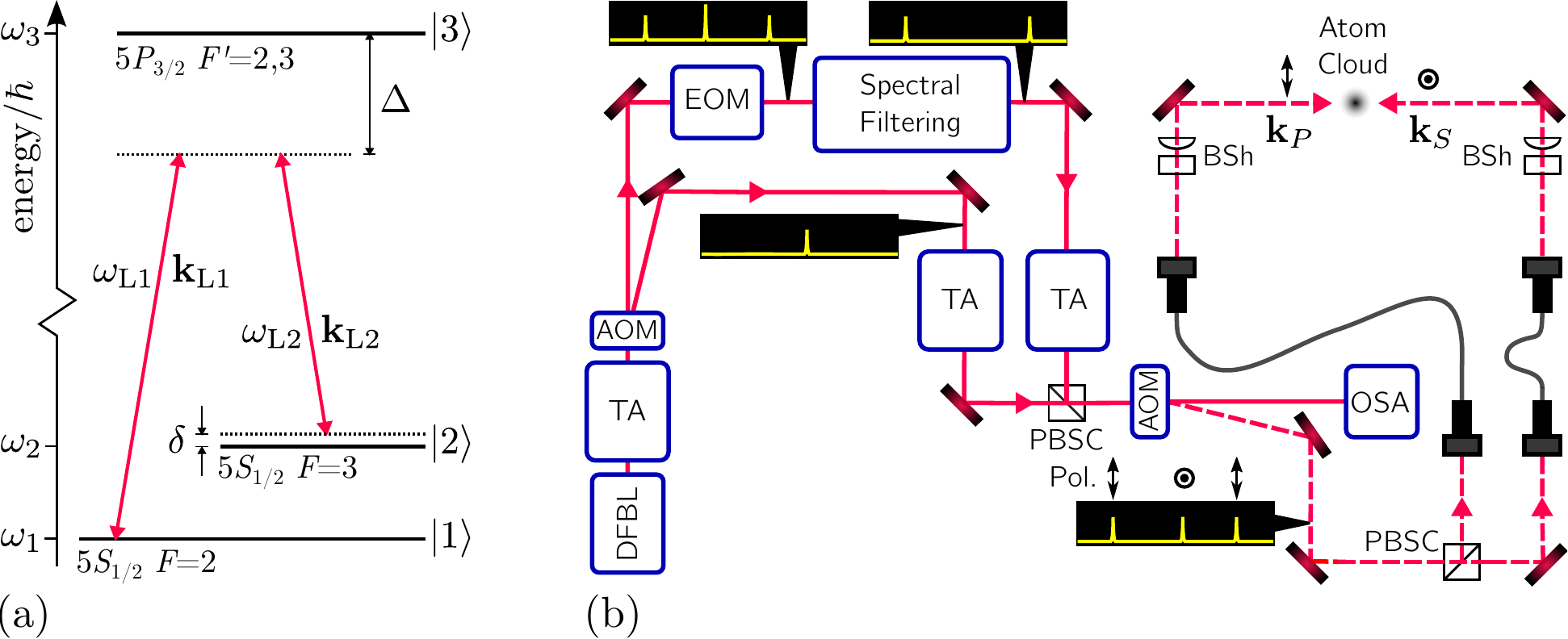}
\caption{(a) Energy-level diagram for the interferometric velocimetry experiment in $^{85}$Rb. (b) Schematic of the experimental setup of the Raman beams: distributed feedback diode laser (DFBL), tapered amplifier (TA), polarizing beam-splitter cube (PBSC), optical spectrum analyzer (OSA), beam shaper and focusing lens (BSh). The annotation bubbles show sketches of the beam spectrum at each preparation stage.}
\label{Experiment}
\end{figure}

Since sensitivity to inertial motion requires that the beamsplitter and recombiner interactions impart an impulse, they must involve optical rather than microwave transitions. For the necessary phase coherence between these interactions, and to use states whose lifetimes do not limit the interferomeric measurement, it is common to use the pseudo-two-level system offered by a two-photon Raman transition~\cite{Bateman07}. Our $\upi/2$ interferometer pulses are realized by driving stimulated Raman transitions between the $5S_{1/2}\,F=2$ and $F=3$ ground hyperfine levels, using 780~nm beams detuned from the $5P_{3/2}$ states, as illustrated in Figure~\ref{Experiment}(a), with the Raman detuning $\delta=0$. Atomic velocities may be measured conventionally by Raman velocimetry, using a long, weak Raman pulse to excite a small velocity class defined by the probe pulse detuning $\delta$~\cite{Kasevich91} and repeating over a range of $\delta$ to derive the velocity distribution.

The source of our Raman pulses is shown schematically in Figure~\ref{Experiment}(b). The continuous-wave beam from a 780~nm distributed feedback diode laser red-detuned from single-photon resonance by $\Delta\approx2\upi\times13$\,GHz is spatially divided by a 310~MHz acousto-optical modulator (AOM), and the rest of the microwave frequency shift is achieved by passing the undeflected beam through a 2.726~GHz electro-optical modulator (EOM). We control the EOM phase and frequency using an in-phase and quadrature-phase (IQ) modulator fed from a pair of arbitrary waveform generators. The carrier wave is removed after the EOM using a stabilized fibre-optic Mach-Zehnder interferometer~\cite{Cooper13} leaving two sidebands, one of which is non-resonant.

The two beams are individually amplified by tapered laser diodes, recombined with orthogonal polarizations and passed through an AOM (rise time $\sim\!100$~ns), whose first-order output forms the Raman pulse beams. These are then separated by a polarizing beam-splitter and passed via optical fibres to the MOT chamber.

After the fibres, each beam is passed through a Topag GTH-4-2.2 refractive beam shaper and 750~mm focal length lens to produce an approximately uniform $1.4$~mm square beam whose intensity varies by $\sim\!15$\% across the MOT cloud. The 310~MHz shifted beam has an optical power of 100~mW and the beam containing the two EOM sidebands has 200~mW. This gives an intensity around $5\, \mathrm{W\,cm}^{-2}$ -- significantly higher than the large-waist Gaussian beams required for the same spatial homogeneity. To avoid broadening effects due to sublevel-dependent light shifts, the Raman beams have orthogonal linear polarizations. Although the phase profile of the top-hat beam is non-uniform~\cite{Boutu11}, an individual atom should not traverse a significant phase gradient during a few-$\mu$s pulse sequence.

\section{Results}

\begin{figure}
\centering
\includegraphics[width=100mm]{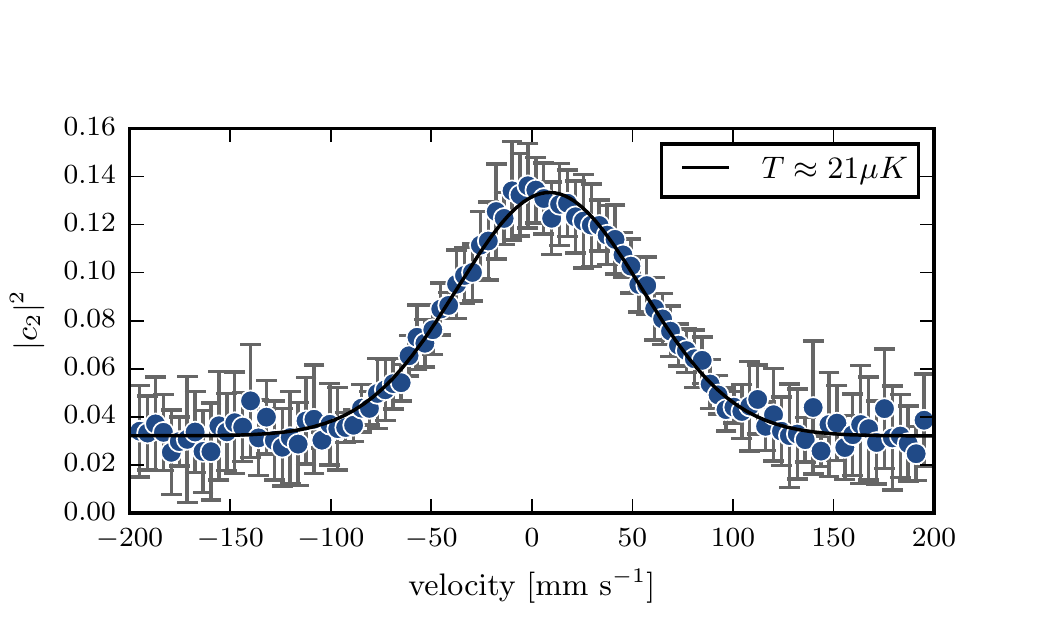}
\caption{Velocity distribution measured by low intensity Raman velocimetry. A thermal distribution with a temperature of $21\,\umu$K sits atop a broad background attributed to inhomogeneous sub-Doppler cooling~\cite{Dunning15,Townsend95}.}
\label{thermal}
\end{figure}

The distribution of velocities within our atom cloud, measured by Raman velocimetry (2.5~mW per Raman beam for 100~$\umu$s), is shown in Figure~\ref{thermal}, and fits well a Gaussian distribution with a temperature of $21\,\umu$K superimposed upon a broad background that we have previously attributed to inhomogeneous sub-Doppler cooling~\cite{Dunning15,Townsend95}. This velocimetry method is itself subject to inaccuracies, and at longer exposure times yields higher temperatures, perhaps for the reasons discussed in~\cite{Hughes17}.

\begin{figure}
\centering
\includegraphics[width=100mm]{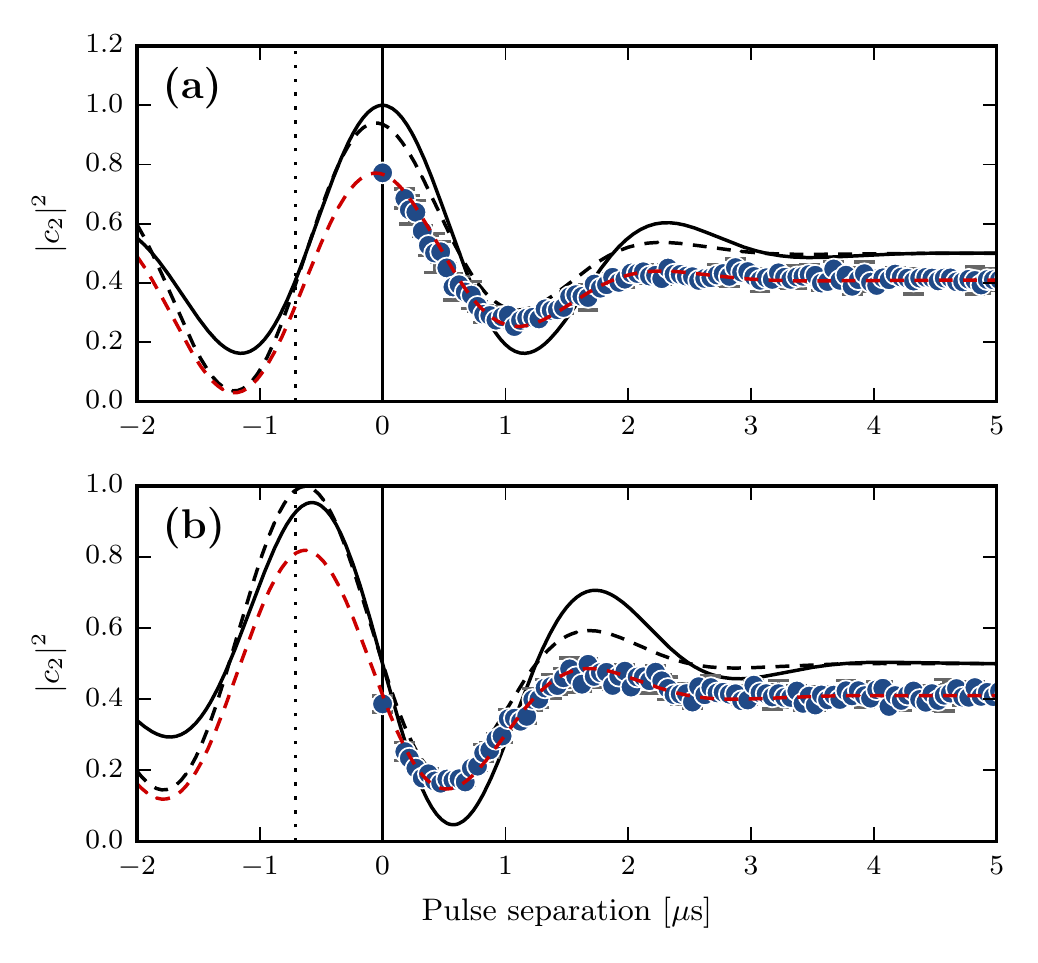}
\caption{(a) In-phase $\mathcal{C}(T)$ and (b) quadrature $\mathcal{S}(T)$ components of the interferometer signal for a Rabi frequency of $450\,\mathrm{kHz}$ and detuning of $400\,\mathrm{kHz}$ to cancel the light shift during the interferometer pulses. Circles show experimental data; solid black curves show predictions assuming perfect beamsplitter pulses; dashed black curves are predictions including the phase and amplitude corrections described in the Appendix; dashed red curves are further scaled by a factor of 0.82, which we attribute to atom loss from the Raman beams~\cite{Dunning14}. The dotted lines indicate the effective time origin $T = -2/\Omega$.}
\label{traces}
\end{figure}

Figure~\ref{traces} shows the in-phase and quadrature interferometer traces, $\mathcal{C}(T)$ and $\mathcal{S}(T)$, for our atom cloud. By initially adjusting the Raman detuning $\delta$ to maximize the population transferred by a $\upi$-pulse, we cancel the light shift during the interferometer pulses but incur a detuning of $400\,\mathrm{kHz}$ in between, giving the traces the form of damped oscillations. The Raman Rabi frequency $\Omega/2\upi$ is around $450\,$kHz.

Figure~\ref{traces} also shows theoretical predictions, assuming the measured velocity distribution, for three scenarios. The solid black curves are for ideal beamsplitter pulses that introduce no phase or amplitude perturbations and correspond to the regime of high Rabi frequency: these show the ideal signal $\vert c_{2} \vert^{2} = 1$ at $T=0$, and perfect symmetry or antisymmetry about this point. The black dashed curves take into account the phase and amplitude corrections resulting from the Doppler sensitivity described in the Appendix. For the red dashed curves, these are then fitted to the experimental results by introducing an empirical scaling factor of 0.82, which accounts for the effect of atoms being lost from the region illuminated by the Raman beams but remaining within the cross-section of the read-out beams~\cite{Dunning14}. The dotted vertical lines show the effective time origin $T = -2/\Omega$. Simulations for the hypothetical region $-2/\Omega < T < 0$ do not account for the light shift due to the a.c. Stark effect, which would cancel the 400~kHz detuning in this region.

\begin{figure}
\centering
\includegraphics[width=100mm]{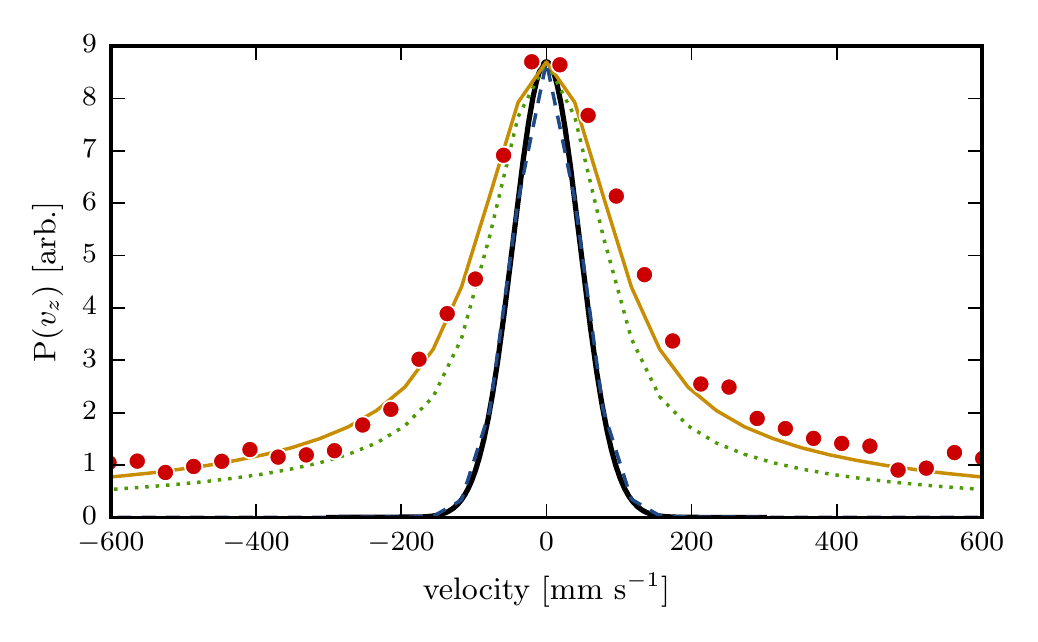}
\caption{The velocity distribution derived by Fourier transformation of interferometric measurements (red circles), corrected for the 400~kHz detuning from the known hyperfine frequency, is rather broader than the $21\,\umu$K distribution measured by Raman velocimetry (solid black curve) because of the residual Doppler sensitivity in the beamsplitter pulses and temporal truncation of the interferometer traces. For comparison, the dashed blue and green curves show the real part and absolute magnitude of the distribution derived by Fourier transformation of the simulated signal for a Gaussian distribution of the same temperature assuming ideal $\upi/2$ pulses, while the yellow curve takes into account the factors described in the Appendix.}
\label{distributions}
\end{figure}

The velocity distribution obtained by Fourier transforming the curves of Figure~\ref{traces} is shown (red circles) in Figure~\ref{distributions}, along with the $21\,\umu$K Gaussian (black solid curve) measured by Raman velocimetry. The dashed curves show simulated results, assuming measurements limited to $0<T<5\,\umu\mathrm{s}$, corresponding to the real part (blue) and magnitude (green) of the derived distribution with ideal beamsplitter pulses, and the solid yellow curve shows the magnitude taking into account the Doppler sensitivity of the beamsplitter and recombiner interactions. Our experimentally-derived distribution shows excellent agreement with the simulation for the latter case.

\section{Conclusion}
The measurement of velocity distributions and the translational temperatures of
cold atom clouds is generally performed using single-photon~\cite{Westbrook90}
or Raman~\cite{Kasevich91} Doppler spectroscopy, recoil-induced
resonances~\cite{Courtois94}, time-of-flight expansion imaging~\cite{Lett88}, or
determination of the release-and-recapture efficiency~\cite{Chu85}. Each
technique has its shortcomings: off-resonant excitation~\cite{Hughes17} and
optical pumping can perturb both the velocity distribution and its measurement;
time-of-flight techniques require either a point-like initial sample or careful
deconvolution; and the highest resolution often incurs a signal-to-noise
penalty, if for example measurements are restricted to a thin imaging region.
 
Atom interferometry offers an alternative method of velocity measurement in
which the atoms are unperturbed between the beamsplitter and recombiner pulses,
whose effects upon the atomic populations and velocities are well defined. We
have demonstrated the use of interferometric velocimetry to measure the
temperature of an ultracold gas of $^{85}$Rb, which by low intensity Raman
velocimetry we determine to be $21\,\umu$K. Accurate interpretation of the
results depends upon good knowledge of the phases and accompanying timing
offsets introduced by residual Doppler effects in the interferometer
interactions. The technique is most suited to the lowest temperatures, and hence
the longest interferometer times $T$, for which the perturbations due to
residual Doppler effects in the $\upi$-pulses have the least effect. Such
long-period measurements are in principle limited only by the residence times of
the expanding cloud within the interferometer beams, and intensity or field
inhomogeneities.

The sensitivity of atom interferometry to the atomic velocities is the basis for atom interferometric measurement of accelerations, rotations, gravitational fields and their gradients, all of which are based upon the differential measurement of velocities in a back-to-back pair of velocity-sensing Ramsey interferometers which, from discrete measurements of the atomic velocity components, reveal the linear or Coriolis accelerations of atoms relative to the apparatus. The interferometer pair in each case allows the interfering paths to be closed, cancelling the path separation phase of Equation~(\ref{phase4}).

%\newpage
\section*{In memory of Prof. Danny Segal}
Our dearly missed friend and colleague Danny Segal was an inspiring physicist, talented musician and artist, industrious handiman and top-notch human being. Mildly spoken and patient, Danny put great thought and imagination into his teaching and exuded a contagious delight in his subject that was picked up by countless students, for whom he cared deeply and whose frailty had his sympathy even when they were in trouble. Noble and kindly, Danny was a source of gentle but profound wisdom, given simply but endlessly recalled. As a researcher he was painstakingly thorough, always intent upon conveying clear insights into complex phenomena. A scrupulously fair referee, he sold his own research on its straight, unembellished merits, and never seemed to suffer for such honesty.
He was joyful company, humorous and observant, often drawing upon his passion for music and love of art, both of which he practiced masterfully. Whether in the lab, playing the blues, building walls or striding the countryside, Danny showed energy and dedication yet had ambition only for his work, pursuits and family.

Danny's research spanned many aspects and regimes of quantum and atomic physics,
from PhD studies of atomic collision dynamics using pulsed dye lasers to his
ultimate expertise with narrow clock transitions of single trapped ions in their
motional ground states~\cite{Goodwin16}. Many topics involved
velocity-dependent interactions of atoms or ions with laser
light~\cite{Thompson97}. With us, he worked on the amplification of
Doppler cooling techniques~\cite{Freegarde06} and the use of Doppler
interferometry as the basis for a quantum computer~\cite{Freegarde03} to be used
not for calculation but for the physical effect upon the atoms comprising it. He
applied similar laser cooling techniques to trapped ions~\cite{Segal14} ---
first in the Doppler regime, addressed here, in which the instantaneous velocity
of the oscillating ion moves the optical interaction locally into and out of
resonance~\cite{Koo04}; and subsequently in the
resolved-sideband Lamb-Dicke regime, in which the optical interaction
adiabatically transforms the entire trapped ion wavefunction from one harmonic
oscillator state to another~\cite{Goodwin16}.

We hope that Danny would have enjoyed the work presented here. We dedicate this paper to him, and remember him with great fondness.

\section*{Acknowledgements}
This work was supported by the EPSRC through the UK Quantum Technology Hub for
Sensors \& Metrology under grant EP/M013294/1, and by Dstl under contracts
DSTLX-1000091758 and DSTLX-1000103157R. The data presented in this paper are
available for download from {\bf https://doi.org/10.5258/SOTON/D0278}.

\bibliographystyle{tfp}
\bibliography{velocimetry,danny}

\appendix

\section{Doppler sensitivity within long $\upi/2$-pulses}

In the Bloch sphere representation~\cite{Feynman57}, detuned Rabi oscillations correspond to rotation with Rabi frequency $\omega \equiv \vert \bm{\uomega} \vert$ about the field vector $\bm{\uomega} \equiv \mathbf{\Omega} + \mathbf{\Delta}$, where $\mathbf{\Omega}$ represents the rotation axis in the equatorial plane when the optical field is resonant and $\mathbf{\Delta}$, parallel to the polar axis, accounts for the detuning $\Delta \equiv \vert \mathbf{\Delta} \vert$. For a given Rabi frequency $\vert \mathbf{\Omega} \vert$, the speed traced out on the surface of the Bloch sphere is independent of $\mathrm{\Delta}$. The trajectory on the Bloch sphere may hence be written as
\begin{eqnarray}
\mathbf{r}(t) & = & \left( \mathbf{r}_{0} \cdot \bm{\hat{\uomega}} \right) \bm{\hat{\uomega}} + \cos \omega t \left[ \mathbf{r}_{0} - \left( \mathbf{r}_{0} \cdot \bm{\hat{\uomega}} \right) \bm{\hat{\uomega}} \right] + \sin \omega t \left( \bm{\hat{\uomega} \times} \left[ \mathbf{r}_{0} - \left( \mathbf{r}_{0} \cdot \bm{\hat{\uomega}} \right) \bm{\hat{\uomega}} \right] \right) \nonumber \\
& = & \left( \mathbf{r}_{0} \cdot \bm{\hat{\uomega}} \right) \bm{\hat{\uomega}} + \cos \omega t \left[ \mathbf{r}_{0} - \left( \mathbf{r}_{0} \cdot \bm{\hat{\uomega}} \right) \bm{\hat{\uomega}} \right] + \sin \omega t \left( \bm{\hat{\uomega} \times} \mathbf{r}_{0} \right),
\end{eqnarray}
where $\bm{\hat{\uomega}}$ is a unit vector in the direction of $\bm{\hat{\uomega}}$ etc. and $\mathbf{r}_{0}$ is the start of the trajectory.

If the atom begins in a pure state, we have $\mathbf{r}_{0}\cdot\mathbf{\Omega} = 0$, $\mathbf{r}_{0}\cdot\mathbf{\Delta} = \Delta$, we find
\begin{eqnarray}
\mathbf{r}(t) & = & \frac{\delta (\mathbf{\Omega} + \mathbf{\Delta})}{\mathrm{\Omega}^{2} + \mathrm{\Delta}^{2}} + \cos \omega t \left[ \mathbf{r}_{0} -  \frac{\Delta (\mathbf{\Omega} + \mathbf{\Delta})}{\mathrm{\Omega}^{2} + \mathrm{\Delta}^{2}} \right] + \sin \omega t \frac{\mathbf{\Omega \times r}_{0}}{\sqrt{\mathrm{\Omega}^{2} + \mathrm{\Delta}^{2}}} \nonumber \\
& = & \mathbf{r}_{0} \left[ \frac{\Delta^{2}}{\Omega^{2} + \delta^{2}} + \cos \omega t \left( 1 -  \frac{\Delta^{2}}{\mathrm{\Omega}^{2} + \mathrm{\Delta}^{2}} \right) \right] + \bm{\mathrm{\Omega}} \left( \frac{\Delta}{\Omega^{2} + \Delta^{2}} - \cos \omega t \frac{\Delta}{\Omega^{2} + \Delta^{2}} \right) \nonumber \\
&& \hspace{5mm} + \frac{\bm{\hat{\Omega} \times \mathrm{r}}_{0}}{\sqrt{\Omega^{2} + \Delta^{2}}} \sin \omega t \nonumber \\
& = & \frac{\Delta^{2} + \Omega^{2} \cos \omega t}{\Omega^{2} + \Delta^{2}} \mathbf{r}_{0} + \left( 1 - \cos \omega t \right) \frac{\Delta \Omega}{\Omega^{2} + \Delta^{2}} \bm{\hat{\Omega}} + \frac{\Omega}{\sqrt{\Omega^{2} + \Delta^{2}}} \bm{\hat{\Omega} \times \mathrm{r}}_{0} \sin \omega t \nonumber \\
& \equiv & \sin \alpha \mathbf{r}_{0} + \cos \alpha \sin \phi \, \bm{\hat{\Omega}} + \cos \alpha \cos \phi \, \bm{\hat{\Omega} \times \mathrm{r}}_{0} ,
\end{eqnarray}
where $\alpha$ and $\phi$ are the latitude and longitude on the Bloch sphere. If the pulse duration is set to provide a rotation of $\upi/2$ on resonance, so that $\Omega t = \upi/2$ then, to lowest order in $\Delta/\Omega$, we find that the latitude and longitude -- which in the ideal case will both be zero -- will be
\begin{eqnarray}
\alpha & = & \sin^{-1} \frac{\Delta^{2} + \Omega^{2} \cos \omega t}{\Delta^{2} + \Omega^{2}}  \approx \left( 1-\frac{\upi}{4} \right) \left( \frac{\Delta}{\Omega} \right)^{2} \label{alpha} \\
\phi & = & \tan^{-1} \frac{\Delta \tan \frac{\omega t}{2}}{\sqrt{\Omega^{2} + \Delta^{2}}} \approx \frac{\Delta}{\Omega} .
\end{eqnarray}
Rotation around the inclined field vector $\bm{\uomega}$ may be decomposed into alternating rotations around $\bm{\Omega}$ and $\bm{\Delta}$, where the latter, corresponding to the free evolution phase for the same period, are offset in part by the azimuthal components of the former, so that the rate at which the longitude varies increases from an initial rate of $\Delta/2$ to a final rate of $\Delta$, averaging $2\Delta/\upi$.

If the interferometer then accrues a free evolution phase $\varphi = \mathbf{k} \cdot \mathbf{v} T$ according to Equation~(\ref{phase5}), followed by a recombiner interaction with the same duration and detuning as calculated here, we obtain sinusoidal fringes that differ from the resonant case by a phase shift whose leading terms are
\begin{equation}
\beta \approx 2 \left( \frac{\Delta}{\Omega} \right) - \frac{10-3\upi}{6} \left( \frac{\Delta}{\Omega} \right)^{3}
\label{phase6}
\end{equation}
and whose amplitude is multiplied by a factor
\begin{equation}
\gamma \approx 1 - \frac{(4-\upi)^{2}}{16} \left( \frac{\Delta}{\Omega} \right)^{4}.
\label{gamma}
\end{equation}
The population transferred oscillates between 0 and $1-2(\upi\!-\!4)^{2}/16 (\Delta/\Omega)^{4}$.

The detuning of the $\upi/2$ pulses typically comprises the velocity-dependent Doppler shift $\Delta_{\mathrm{Doppler}} = \mathbf{v} \cdot \mathbf{k}$ and a constant $\Delta_{\mathrm{light}}$ accounting for the light shift and any other steady offset. The combined effect %of the intended velocity dependence of $\varphi$ and the leading term in $\beta$
is, to leading order in Equation~(\ref{phase6}), an apparent offset in the interferometer period $T$ and a shift in the apparent velocity:
\begin{eqnarray}
\varphi_{\mathrm{total}} & = & \mathbf{v} \cdot \mathbf{k} T + 2 \left( \frac{\mathbf{v} \cdot \mathbf{k}}{\Omega} \right) + \Delta_{\mathrm{light}}T \nonumber \\
& = & \left( \mathbf{v} \cdot \mathbf{k} + \Delta_{\mathrm{light}} \right) \left( T + \frac{2}{\Omega} \right) - \frac{2\Delta_{\mathrm{light}}}{\Omega} .
\end{eqnarray}
The constant final term and the offset in $T$ merely introduce a complex phase -- constant and velocity-dependent respectively -- into the derived velocity component, without changing its amplitude. The velocity shift displaces the derived velocity distribution. Higher order terms could broaden or distort further the derived velocity distribution.

More significantly, the offset in $T$ limits the range of effective interferometer periods that may be explored. Restricting measurements to $T>\tau$ is equivalent to multiplying the interferometer traces by the Heaviside function $\mathcal{H}(t\!-\!\tau)$. However, as long as the underlying interferometer pattern is symmetrical about $T=0$, its Fourier transform is real and, for $\tau=0$,  convolution with the transform of the Heaviside function
\begin{equation}
\mathrm{FT}\left\{ \mathcal{H}(t\!-\!\tau) \right\} = \frac{1}{\sqrt{2\upi} \, \omega}\mathrm{i} \exp \mathrm{i}\omega\tau + \sqrt{\frac{\upi}{2}} \delta(\omega)
\end{equation}
has no effect upon the real component of the derived distribution, although it introduces an imaginary term that broadens the derived distribution if the magnitude rather than the real part is used to determine it.

If the Heaviside function is displaced by the effective offset $\tau = 2/\Omega$, this is no longer true,
%and the phase shift also leads to uncertainty about the expected quadrature of the distribution. The
and the magnitude of the derived distribution is again broader than the actual velocity distribution. For a Gaussian distribution $\rho(v_{\mathrm{k}}) \propto \exp [ - ( v_{\mathrm{k}}/\Delta v )^{2} ]$, which if $\Delta v = \sqrt{k_{\mathrm{B}}\Theta/m}$ represents the thermal distribution for atoms of mass $m$ at a temperature $\Theta$, the effect is to enhance the wings of the distribution by multiplication with the complementary error function
\begin{equation}
1 - \erf \left( \tau + \mathrm{i} \frac{v_{\mathrm{k}}}{\Delta v} \right).
\end{equation}

We note that composite pulse techniques~\cite{Dunning14a} could allow reduction of these systematic perturbations to the interferometer phase.

\end{document}